\documentclass[11pt,a4paper]{article}
\pdfoutput=1
\usepackage{jheppub}

\usepackage{amsmath}
\usepackage{verbatim}
\usepackage{graphicx}
\usepackage{mathrsfs}
\usepackage{appendix}
\usepackage{caption}
\usepackage{float}
\usepackage{subfigure}
\usepackage{epstopdf}
\usepackage{multicol}
\usepackage{tikz}

\newcommand{\e}{\epsilon}

\newcommand{\be}[1]{ \begin{equation}\label{#1} }
	\newcommand{\ee}{\end{equation}}
\newcommand{\bea}[1]{\begin{eqnarray}\label{#1} }
	\newcommand{\eea}{\end{eqnarray}}
\newcommand{\bes}{\begin{subequations}}
	\newcommand{\ees}{\end{subequations}}

\newcommand{\p}{\partial}

\newcommand{\non}{\nonumber}

\newcommand{\ie}{\emph{i.e.}}

\title{BRST Symmetry of Non-Lorentzian Yang-Mills Theory}

\author{Minhajul Islam} \author{\\}

\affiliation{Indian Institute of Technology Kanpur, Kalyanpur, Kanpur 208016. INDIA. \\}

\emailAdd{ minhajul@iitk.ac.in}

\abstract{ 
	We explore the realization of BRST symmetry in the non-Lorentzian Yang-Mills Lagrangian within the context of Galilean and Carrollian Yang-Mills theory. Firstly we demonstrate the nilpotent property of classical BRST transformations and construct corresponding conserve charges for both cases. Then we analyze the algebra of these charges and observe the nilpotent properties at the algebraic level. The findings of this study contribute to a deeper understanding of BRST symmetry in non-Lorentzian Yang-Mills Lagrangians and provide insights into the algebraic properties of related conserve charges.
}

\begin{document}
	
	\maketitle

	\section{Introduction}
	
Yang-Mills theories, named after Chen Ning Yang and Robert Mills \cite{Yang:1954vj}, are quantum field theories (QFT) that describe the behavior of elementary gauge bosons. This is the main ingredient of the Standard Model of particle physics, which is our best current understanding of the behavior of subatomic particles. 
	
\medskip
	
Our current comprehension of the physical world thus heavily relies on the framework of QFT. The textbook formulation of QFT is closely tied to the principles of relativistic physics, including Lorentz and Poincar{\'e} symmetry. However, when examining real-life systems, it is often necessary to consider approximations and limits of the fundamental theory. In this paper, we will be interested in QFTs where Poincar{\'e} symmetry is replaced by Galilean and Carrollian symmetries which constitute the the low and high energy sectors of relativistic QFTs.

\medskip
	
To understand Galilean and Carrollian QFTs, we will adopt a group-theoretic approach starting from the Poincaré algebra and taking the limits of large c (speed of light) and small c. These limits yield two different symmetry algebras: the familiar Galilean algebra and the less familiar Carrollian algebra. In both limits, several counter-intuitive concepts emerge. The spacetime metrics degenerate, light-cones open up for non-relativistic theory, and close up for Carrollian theory. Moreover, symmetry algebra gets enhanced in both cases.
	
	\medskip
	
	Galilean theories, which correspond to the limit of $c\rightarrow \infty$ (where c is the speed of light), are important for various areas of physics such as condensed matter physics, non-AdS holography, and hydrodynamics. In this limit, the metric of spacetime degenerates, and the structure of spacetime changes from the usual Riemannian structure to a new one called Newton-Cartan spacetime \cite{Duval:1984cj,Duval:2009vt,Bleeken:2015ykr,Bergshoeff:2015sic,Hansen:2020pqs}.
	For developing non-relativistic physics involves starting from a Poincaré-invariant theory and expanding it using a large c-expansion provides many insights into non-relativistic physics, such as enhanced symmetry algebra and actions at each order \cite{Hansen:2019svu,Hansen:2020wqw, Hansen:2020pqs, Ergen:2020yop}.

	\medskip    
	
	The Carrollian limit is the opposite of the limit mentioned earlier, corresponding to $c\rightarrow 0$. The Carroll algebra was first discussed in \cite{LevyLeblond,NDS}. It has recently become important in various applications, particularly in the understanding of flat space holography \cite{Susskind:1998vk}. AdS/CFT duality is one of the most promising tools for understanding quantum gravity. When the radius of curvature is infinite, AdS spacetime becomes flat spacetime. Correspondingly, on the dual side, sending the speed of light to zero results in a Carrollian conformal field theory \cite{Bagchi:2012cy}.  Some important references for holography for asymptotically flat spacetime are \cite{Susskind:1998vk, Barnich:2010eb, Bagchi:2010eg,Bagchi:2012cy, Bagchi:2014iea,Bagchi:2016bcd,Barnich:2012aw,Barnich:2012xq}. The understanding of flat space holography recently has taken two different directions, viz. Celestial holography and Carrollian holography. Celestial holography relates gravity in 4d asymptotically flat spacetimes to a 2d CFT living on the celestial sphere \cite{Raclariu:2021zjz,Pasterski:2021rjz}. On the other hand, Carrollian holography relates 4d asymptotically flat gravity to 3d Carrollian CFTs living on the entire null boundary of 4d bulk spacetime \cite{Bagchi:2016bcd,Bagchi:2019xfx,Bagchi:2019clu,Duval:2014uva,Duval:2014lpa, Dutta:2022vkg, Bagchi:2023fbj, Nguyen:2023vfz, Salzer:2023jqv, Saha:2023hsl}. Some works \cite{Bagchi:2022emh,Donnay:2022aba, Donnay:2022wvx} connect both formalisms.
	
\medskip
	
Carrollian physics appears on any null hypersurface, including the horizon of a black hole \cite{Donnay:2019jiz, Bicak:2023vxs}. Carrollian gravity may provide a tractable version of general relativity and may be useful in various physical contexts \cite{Hartong:2015xda,Bergshoeff:2017btm}. The theory of Carroll is also important in cosmology, inflation \cite{deBoer:2021jej}, 
for fluids flowing at very high velocities \cite{Bagchi:2023ysc}, fractons \cite{Bidussi:2021nmp,Perez:2022kax, Figueroa-OFarrill:2023vbj}, and the study of flat physics in condensed matter systems \cite{Bagchi:2022eui}.The Carrollian limit of the string theory worldsheet leads to the very high energy tensionless regime of strings  \cite{Bagchi:2013bga,Bagchi:2015nca,Bagchi:2020fpr}.

\medskip

Before we begin our investigations of aspects of non-Lorentzian QFTs, we will quickly summarize summarize previous research on Galilean  and Carrollian gauge theories. Galilean electrodynamics was first studied a long time ago in \cite{LBLL}. Later, in \cite{Bagchi:2014ysa,Bagchi:2015qcw,Bagchi:2017yvj}, authors discovered infinite-dimensional Galilean conformal symmetry in both Galilean abelian and Galilean Yang-Mills theory at the level of equations of motion. More detailed work has been done on constructing actions for both Galilean abelian \cite{Festuccia:2016caf,Banerjee:2019axy} and Yang-Mills theory \cite{Bagchi:2022twx}. The quantum properties of Galilean scalar electrodynamics were examined in \cite{Chapman:2020vtn},  Galilean QED, Scalar QED, non-linear electrodynamics in \cite{Banerjee:2022uqj, Baiguera:2022cbp, Sharma:2023chs, Mehra:2023rmm,Banerjee:2022sza}.
	
		\medskip
	
In the Carroll case, conformal structures at the level of equations of motion in \cite{Bagchi:2016bcd,Bagchi:2019clu,Bagchi:2019xfx}. \cite{Basu:2018dub} presented the Carrollian action for the so-called electric abelian theory, which is an interacting field theory with a scalar field \cite{Banerjee:2020qjj,deBoer:2021jej}. Using the small c-expansion, the magnetic sector of Carrollian abelian theory has been recently constructed in \cite{deBoer:2021jej}, and the conformal structure of this magnetic action was analyzed. In \cite{Henneaux:2021yzg}, the authors constructed the off-shell Carrollian Yang-Mills theory in the Hamiltonian formulation. Finally, in \cite{Islam:2023rnc}, the action formulation for Carrollian Yang-Mills theory was constructed. In \cite{Chen:2023pqf} the authors constructed Carrollian field theory using null reduction techniques.

	\medskip

BRST symmetry, first introduced by Becchi, Rouet, Stora, and Tyutin in the 1970s \cite{Becchi:1975nq,Becchi:1974md}, plays a crucial role in in the study of gauge theories. BRST symmetry is a type of ghost symmetry, which means that it involves the introduction of additional degrees of freedom that are not physical, but help to resolve the ambiguity in the system. BRST symmetry of $SU(N)$ Yang-Mills theory is defined using a set of operators known as BRST operators, which act on the fields of the system and generate BRST symmetry transformations. BRST operators have the property that they are nilpotent, which means that they square to zero. This property is crucial in the construction of the BRST symmetry, as it ensures that the BRST transformations form a closed algebra. BRST symmetry of $SU(N)$ Yang-Mills theory is important for the computation of physical observables in the system. 
	
In this paper, we will construct BRST symmetry for Galilean and Carrollian Yang-Mills theories. We incorporate important results and notation from our recent previous works, namely  \cite{Bagchi:2022twx} and \cite{Islam:2023rnc}.  We first provide a brief review of these theories in Section \ref{review}. We then proceed to explore BRST symmetry for Galilean field theories in Section \ref{BRSTG}, and in Section \ref{BRSTCarrollian} we do the same for Carrollian field theories. Finally, in Section \ref{Conclusion}, we conclud the paper by summarizing our findings and discussing the implications of our results. Overall, our analysis sheds light on BRST symmetry in non-Lorentzian field theories, and provides a foundation for future investigations in this area.

	\section{Brief review of Non-Lorentzian Yang-Mills}\label{review}
\subsection{Galilean Yang-Mills}\label{GalileanReview}

In \cite{Bagchi:2022twx}, we constructed the action for Galilean Yang-Mills theory by using the null reduction procedure. We will give a brief review on that. We write the Lagrangian density of Yang-Mills theory in $(d+1)$ dimensions in null coordinates:
\begin{eqnarray}\label{hjed}&&
	\hspace{-.9cm}	\mathcal{L}_{YM}= -\frac{1}{4}\eta^{\tilde{\mu}\tilde{\rho}}\eta^{\tilde{\nu}\tilde{\sigma}}F_{\tilde{\mu}\tilde{\nu}}^{a}F_{\tilde{\rho} \tilde{\sigma}}^{a}=-\frac{1}{4}\Big[2F^{a}_{ut}F^{a}_{tu}+F^{ija}F^{a}_{ij}+4F^{a}_{ui}F^{ia}_{t}\Big],
\end{eqnarray}
and perform null reduction along the null direction parametrized by the coordinate $u$. We take the gauge field to be independent of the $u$-coordinate, \ie\ $\partial_u A_{\tilde{\mu}}^a = 0$, and decompose its components as
\begin{eqnarray}\label{frn}
	A_{u}^{a}=\phi^a,\quad A_{t}^a=a_{t}^a,\quad A_{i}^{a}=a_{i}^{a}.
\end{eqnarray}
Then the null reduction gives the Lagrangian density in $d$ spacetime dimensions as
\bea{}&&\hspace{0cm}
\mathcal{L}_{GYM}=\Big[\frac{1}{2}(\partial_{t}\phi^{a}-gf^{abc}\phi^{b}a_{t}^{c})(\partial_{t}\phi^{a}-gf^{ade}\phi^{d}a_{t}^{e})\non\\&&
\hspace{1.5cm}-\frac{1}{4}(\partial^{i}a^{j}-\partial^{j}a^{i}+gf^{ade}a^{id}a^{je})(\partial_{i}a_{j}-\partial_{j}a_{i}+gf^{abc}a_{i}^{b}a_{j}^{c}) \non\\&&
\hspace{1.5cm}
+(\partial_{i}\phi^{a}-gf^{abc}\phi^{b}a_{i}^{c})(\partial_{t}a^{ia}-\partial^{i}a_{t}^{a}+gf^{abc}a_{t}^{b}{a}^{ic})\Big],
\eea
where the subscript GYM stands for Galilean Yang-Mills. It can also be written in a compact form given by 
\begin{eqnarray}\label{ngym}
	\mathcal{L}_{GYM}=\frac{1}{2}D_{t}\phi^{a}D_{t}\phi^{a}+D_{i}\phi^{a}E^{ia}-\frac{1}{4}W^{ija}W_{ij}^{a},
\end{eqnarray}
where $D_t$, $D_i$ are gauge-covariant derivatives and $E^{ia}$, $W^a_{ij}$ are field strength variables defined as
\bes{}\label{quan}
\begin{eqnarray}\label{Covariant form}&&
	D_{t}\phi^{a}=\partial_{t}\phi^{a}-gf^{abc}\phi^{b}a_{t}^{c},\quad
	D_{i}\phi^{a}=\partial_{i}\phi^{a}-gf^{abc}\phi^{b}a_{i}^{c},\\&&
	E^{ia}=\partial_{t}a^{ia}-\partial^{i}a_{t}^{a}+gf^{abc}a_{t}^{b}{a}^{ic},\quad
	W_{ij}^{a}=\partial_{i}a_{j}-\partial_{j}a_{i}+gf^{abc}a_{i}^{b}a_{j}^{c}.
\end{eqnarray}
\ees
The EOM for the Lagrangian \eqref{ngym} are given by
\bes{}\label{EOM of Yang-Mills}
\begin{eqnarray}&&
	D_{t}D_{t}\phi^{a}+D_{i}E^{ia}=0,\label{EOM-1}\\&&
	D_{i}D_{i}\phi^{a}+gf^{abc}\phi^{b}D_{t}\phi^{c}=0,\label{EOM2}\\&&
	D_{t}D_{i}\phi^{a}-D_{j}W_{ji}^{a}-gf^{abc}\phi^{b}E^{ic}=0.\label{EOM3}
\end{eqnarray}\ees
We can also find these equations by doing the procedure of null reduction on relativistic equations.\footnote{The Lagrangian was also introduced in \cite{Gomis:2020fui} to derive the EOM for Galilean Yang-Mills theory with $U(N)$ gauge group obtained as an effective theory from non-relativistic open string theory.}

\medskip

\noindent If we put $\phi^{a}=0$, the equations become
\begin{eqnarray}\label{phi0}
	D_{i}E^{ia}=0,~~
	D_{j}W_{ji}^{a}=0.
\end{eqnarray}
An interesting point to re-emphasise is that the EOM even with the fields $\phi^a$ turned off are different from the ones obtained by taking limits as was described earlier in the section. It is possible that one needs to consider scalings of the gauge coupling $g$ in order to obtain these results{\footnote{A similar phenomenon was observed when constructing the action of Carrollian scalar electrodynamics in \cite{Bagchi:2019clu}}}. 
\subsection{Carrollian Yang-Mills}\label{CarrollianReview}

In \cite{Islam:2023rnc}, we analyzed the Carrollian limit of Yang-Mills theory and obtained electric and magnetic sectors, where one subsector of each contained non-abelian or self-interaction terms while the other subsector contained copies of the Carrollian abelian theory. In the above mentioned paper, we obtained Carrollian Yang-Mills actions by taking a small $c$-expansion of the Poincaré invariant Yang-Mills action, where different values of the parameter $\delta$ lead to different sectors for Carrollian Yang-Mills theory. 

All four sectors were found to be invariant under infinite Carrollian conformal algebra in $4$ dimensions. Below we provide a brief review of the two non-trivial sectors, electric and magnetic, which are relevant to our current purpose.

\subsubsection*{Electric Action}\label{Non-trivial electric}
The electric sector action which has a non-abelian term, can be written in compact form:
\begin{eqnarray}\label{Electric0}&&
	\mathcal{L}_{0}=\frac{1}{2}\bigg((\p_{t}a_{i}^{a(0)}-\p_{i}a_{t}^{a(0)})(\p_{t}a_{i}^{a(0)}-\p_{i}a_{t}^{a(0)})+2gf^{abc}(\p_{t}a_{i}^{a{(0)}}-\p_{i}a_{t}^{a(0)})a_{t}^{b(0)}a_{i}^{c(0)}\non\\&&
	\hspace{5cm}+g^{2}f^{abc}f^{ade}a_{t}^{b(0)}a_{i}^{c(0)}a_{t}^{d(0)}a_{i}^{e(0)}\bigg)=\frac{1}{2}E_{i}^{a(0)}E_{i}^{a(0)},
\end{eqnarray}
where $E_{i}^{a(0)}=\p_{t}a_{i}^{a(0)}-\p_{i}a_{t}^{a(0)}+gf^{abc}a_{t}^{a(0)}a_{i}^{a(0)}$. The equations of motion following from the action are given by
\bes\label{EOM}
\begin{eqnarray}&&
	\p_{i}E_{i}^{a(0)}+gf^{abc}a_{i}^{b(0)}E_{i}^{c(0)}=D_{i}^{(0)}E_{i}^{a(0)}=0\label{0eeom1},\\&&
	\p_{t}E_{i}^{a(0)}+gf^{abc}a_{t}^{b(0)}E_{i}^{c(0)}=D_{t}^{(0)}E_{i}^{a(0)}=0\label{0eeom2},
\end{eqnarray}
\ees
where $D_{i}\mathcal{O}^{a}=\p_{i}\mathcal{O}^{a}+gf^{abc}a_{i}^{b(0)}\mathcal{O}^{c},\, D_{t}\mathcal{O}^{a}=
\p_{t}\mathcal{O}^{a}+gf^{abc}a_{t}^{b(0)}\mathcal{O}^{c}$.
\medskip

The gauge transformations under which the action \eqref{Electric0} is invariant are given by
\bes\label{electric gauge symmetry}
\begin{eqnarray}&&
	a_{t}^{a(0)}\rightarrow a_{t}^{a(0)'}=a_{t}^{a(0)}+\frac{1}{g}\p_{t}\alpha^{a}+f^{abc}a_{t}^{b(0)}\alpha^{c},\\&&
	a_{i}^{a(0)}\rightarrow a_{i}^{a(0)'}=a_{i}^{a(0)}+\frac{1}{g}\p_{i}\alpha^{a}+f^{abc}a_{i}^{b(0)}\alpha^{c}.
\end{eqnarray}
\ees
This gauge transformation is the same as parent theory, but now we cannot write it in covariant form like relativistic theory. Because, like the non-relativistic theory, the metrics in Carrollian theory are degenerate, and time and space are not on the same footing.

\subsubsection*{Magnetic Action}\label{non-trivial magnetic}

Now, we will discuss on the magnetic sector. The next to leading order( NLO) Lagrangian take from \cite{Islam:2023rnc} contains leading order and NLO fields. 
From the expansion of action section, we have the NLO Lagrangian (coefficient of $c^{0}$) as
\begin{eqnarray}\label{wron0mag}
	\mathcal{L}^{(1)}=\big(D_{t}^{(0)}a_{i}^{a(1)}\big)E_{i}^{a(0)}-\big(D_{i}^{(0)}a_{t}^{a(1)}\big)E_{i}^{a(0)}-\frac{1}{4}f^{ija(0)}f_{ij}^{a(0)}.
\end{eqnarray}
If we take the variation of the Lagrangian with respect to next to leading order fields $a_{t}^{a(1)},a_{i}^{a(1)}$ we will get  Eq.\eqref{EOM}, leading order equations of motion as a property of this formalism. If we take variation with respect to leading order fields ($a_{t}^{a(0)},a_{i}^{a(0)}$), equations of motion are
\bes\label{NLO EOM}
\begin{eqnarray}&&
	D_{i}^{(0)}D_{i}^{(0)}a_{t}^{a(1)}-D_{i}^{(0)}D_{t}^{(0)}a_{i}^{a(1)}-gf^{abc}a_{i}^{b(1)}E_{i}^{c(0)}=0,\\&&
	D_{t}^{(0)}D_{t}^{(0)}a_{i}^{a(1)}-D_{t}^{(0)}D_{i}^{(0)}a_{t}^{a(1)}-gf^{abc}a_{t}^{b(1)}E_{i}^{c(0)}-D_{k}^{(0)}f_{ki}^{a(0)}=0,
\end{eqnarray}
\ees
where $D_{k}^{(0)}f_{ki}^{a(0)}=\p_{k}f_{ki}^{a(0)}+gf^{abc}a_{k}^{b(0)}f_{ki}^{c(0)}$.
Although the action and the equations of motion look nice in compact form, these are not Carroll invariant. To make Carroll invariant, we have to take the constraint $E_{i}^{a(0)}=0$ at the level of action Eq.\eqref{wron0mag}. Then action will become  $-\frac{1}{4}f^{ija(0)}f_{ij}^{a(0)}$ and equations of motion will be $D_{k}^{(0)}f_{ki}^{a(0)}=0$. 

We can derive the Carroll invariant magnetic sector from the relativistic Yang-Mills action if we consider a Lagrange multiplier in relativistic Lagrangian and then take speed of light to zero limit. The relativistic Lagrangian with Lagrange multiplier $\xi_{i}^{a}$ and explicit $c$ factor is given by
\begin{eqnarray}
	\mathcal{L}=-\frac{c^{2}}{2}\xi_{i}^{a}\xi_{i}^{a}+\xi_{i}^{a}F_{0i}^{a}-\frac{1}{4}F_{ij}^{a}F_{ij}^{a}.
\end{eqnarray}
From here, we can get back to the usual Yang-Mills action if we integrate out $\xi_{i}$ fields. Now we can see if we take the small $c$ limit here, we will get
\begin{eqnarray}\label{omag}&&
	\mathcal{L}^{NLO}=\xi_{i}^{a}(\p_{t}a_{i}^{a(0)}-\p_{i}a_{t}^{a(0)})-
	\frac{1}{4}(\p_{i}a_{j}^{a}-\p_{j}a_{i}^{a})(\p_{i}a_{j}^{a}-\p_{j}a_{i}^{a})+
	gf^{abc}a_{t}^{b}a_{i}^{c}\xi_{i}^{a}\non\\&&
	\hspace{2cm}-gf^{abc}a_{i}^{b}a_{j}^{c}\p_{i}a_{j}^{a}-\frac{1}{4}g^{2}f^{abc}f^{ade}a_{i}^{b}a_{j}^{c}a_{i}^{d}a_{j}^{e}
	=\xi_{i}^{a}E_{i}^{a}-\frac{1}{4}f_{ij}^{a}f_{ij}^{a}.
\end{eqnarray}
The Lagrangian contains non-trivial self-interaction terms or non-abelian terms.  The equations of motion of this action are
\begin{eqnarray}\label{0meom}&&
	E_{i}^{a}=0,\quad D_{i}\xi_{i}^{a}=0,\quad D_{t}\xi_{i}-D_{j}f_{ji}=0.
\end{eqnarray}
Here we are getting the constraints $E_{i}^{a(0)}=0$ as an equations of motion for the Lagrange($\xi_{i}^{a}$). Below we will see the full spacetime symmetry of this action.

\medskip

The action Eq.\eqref{omag} is invariant under the gauge transformation
\bes
\begin{eqnarray}&&
	a_{t}^{a}\rightarrow a_{t}^{'a}=a_{t}^{a}+\frac{1}{g}\p_{t}\alpha^{a}+f^{abc}a_{t}^{b}\alpha^{c},\\&&
	a_{i}^{a}\rightarrow a_{i}^{'a}=a_{i}^{a}+\frac{1}{g}\p_{i}\alpha^{a}+f^{abc}a_{i}^{b}\alpha^{c},\\&&
	\xi_{i}^{a}\rightarrow \xi_{i}^{'a}=\xi_{i}^{a}+f^{abc}\xi_{i}^{b}\alpha^{c}.
\end{eqnarray}
\ees
The temporal and spatial component of the gauge field is transformed in the same way as the electric sector. The Lagrange multiplier $\xi_{i}^{a}$ transforms as a scalar in the adjoint representation of the underlying gauge group.

		\section{BRST Symmetry}\label{BRSTNL}
		\subsection{Yang-Mills theory}\label{BRSTG}

	One of the most important techniques for quantization of gauge theories is BRST quantization. The canonical quantization of the modified action obtained after path integral formulation is important to define physical states and {to eliminate the gauge symmetry when restricted to the subspace of the physical Hilbert space}. 
	
	The total gauge-fixed Lagrangian of the relativistic Yang-Mills theory including the gauge fixing and ghost terms is
	\begin{eqnarray}\label{relativisticfulllagrangian}
		\mathcal{L}=-\frac{1}{4}F^{\mu\nu a}F^{a}_{\mu\nu}-\frac{1}{2\xi}\big(\partial^{\mu}A_{\mu}\big)^{2}-\partial^{\mu}\bar{c}^{a}D_{\mu}c^{a},
	\end{eqnarray}
	where $D_{\mu}c^{a}=\partial_{\mu}c^{a}-gf^{abc}A_{\mu}^{b}c^{c}$.
	This gauge fixed Lagrangian is not invariant under the gauge symmetry but invariant under a global symmetry, \emph{i.e.} BRST symmetry. The BRST transformations are given by
	\begin{eqnarray}&&
		\delta A_{\mu}^{a}=\frac{\omega}{g}D_{\mu}c^{a}, \quad
		\delta c^{a}=-\frac{\omega}{2}f^{abc}c^{b}c^{c}, \quad
		\delta \bar{c}^{a}=\frac{\omega}{g\xi}\big(\p_{\mu}A^{\mu a}\big)^{2},
	\end{eqnarray}
	where $\omega$ is an anti-commutating constant parameter. The BRST invariance of the Lagrangian leads to many interesting consequences. It helps us in the covariant quantization of the Yang-Mills theory, to prove unitarity of the S-matrix, {with gauge fixing and ghost term of the Lagrangian}. The BRST invariance also leads to Ward-Takahashi identities; in Yang-Mills theory these are called Slavnov-Taylor identities.
	
	\subsection{Galilean Yang-Mills theory}
	
	Performing the null reduction of the full Lagrangian \eqref{relativisticfulllagrangian}, the NR Yang-Mills Lagrangian with gauge fixing term and ghost term is
	\begin{eqnarray}
		\mathcal{L}_{Gfull}=\frac{1}{2}D_{t}\phi^{a}D_{t}\phi^{a}+D_{i}\phi^{a}E^{ia}-\frac{1}{4}f^{ija}f_{ij}^{a}-\frac{1}{2\xi}(\partial_{t}\phi^{a}+\partial^{i}a_{i}^{a})^{2}-gf^{abc}\partial_{t}\bar{c}^{a}c^{b}\phi^{c}\non\\
		-\partial_{i}\bar{c}^{a}(\partial_{i}c^{a}-gf^{abc}c^{b}a_{i}^{c}).
	\end{eqnarray}
	This Galilean Yang-Mills Lagrangian is invariant under the transformations
	\begin{eqnarray}&&
		\delta\phi^{a}=\omega f^{abc}\phi^{b}c^{c},\,
		\delta a_{t}^{a}=\frac{\omega}{g}D_{t}c^{a},\,
		\delta a_{i}^{a}=\frac{\omega}{g}(D_{i}c^{a}),\non\\&&
		\delta c^{a}=-\frac{\omega}{2}f^{abc}c^{b}c^{c},\,
		\delta \bar{c}^{a}=\frac{\omega}{g\xi}(\partial_{t}\phi^{a}+\partial_{i}a_{i}^{a}) .\label{Galilean-BRST-transformations}
	\end{eqnarray}
	Here $c^{a}$ is a Grassmannian variable and represents the ghost field. The transformations of gauge fields contain ghost field. We can see that the transformations of the gauge fields are actually same as the gauge transformation but now the gauge parameter is replaced by the ghost field $c^{a}$. Thus the Yang-Mills piced of the total Lagrangian is naturally invariant under the above transformations. Under these transformations the gauge-fixing and the ghost terms transform as
	\begin{eqnarray}
		\delta\big(\mathcal{L_{GF+GH}}\big)=gf^{abc}\p_{t}\big[\frac{\omega}{g\xi}\big(\p_{t}\phi^{a}+\p_{i}a_{i}^{a}\big) c^{b}\phi^{c}\big]-\p_{i}\big[\frac{\omega}{g\xi}\big(\p_{t}\phi^{a}+\p_{k}a_{k}^{a}\big)D_{i}c^{a}\big],
	\end{eqnarray}
	which is a total divergence, thus showing invariance of the gauge-fixing and ghost terms.
	
	In relativistic Yang-Mills theories, the BRST transformations are nilpotent, \emph{i.e.} $\delta_1\delta_2\Phi^a = 0$, where $\Phi^a$ denotes all the fields in the theory.
	The nilpotency of the BRST operator is very important to introduce physical states.
	
	Let us check the nilpotency of the BRST transformations \eqref{Galilean-BRST-transformations} in Galilean Yang-Mills theory:
	\bes{}	
	\begin{eqnarray}&&
		\delta^{2}\phi^{a}=-\omega\delta (gf^{abc}c^{b}\phi^{c})=0, \quad \text{using Jacobi identity}\\&&
		\delta^{2}a_{t}^{a}=\frac{\omega}{g} \delta (D_{t}c^{a})=0,~
		\delta^{2}a_{i}^{a}=\frac{\omega}{g}\delta (D_{i}c^{a})=0.
	\end{eqnarray} \ees
	To see this we have to use Jacobi identity for the structure constants
	\begin{eqnarray}
		f^{cab}f^{ckl}+f^{cal}f^{cbk}+f^{cak}f^{clb}=0,
	\end{eqnarray}
	which is derived from the Lie algebra of the underlying gauge group
	\begin{eqnarray}
		[T^{b},[T^{k},T^{l}]]+[T^{k},[T^{l},T^{b}]]+[T^{l},[T^{b},T^{k}]]=0.
	\end{eqnarray}
	For the ghost field, using the Jacobi identity we get
	\begin{eqnarray}
		\delta^{2}c^{a}=\frac{\omega}{2}Q (f^{abc}c^{b}c^{c})=0. 
	\end{eqnarray}
	However for the anti-ghost field, the transformations are nilpotent only upon using the equation of motion for the ghost field, \emph{i.e.}
	\begin{eqnarray}
		\delta^2 \bar{c}^a = \delta (\partial_{t}\phi^{a}+\partial^{i}a_{i}^{a}) = 0
		~~\text{upon using}~~
		\p_{i}D^{i}c^{a}-gf^{abc}\p_{t}(c^{b}\phi^{c})=0.
	\end{eqnarray}
	Now we have seen that the Lagrangian is invariant under the BRST transformations \eqref{Galilean-BRST-transformations} off-shell but to get full nilpotent BRST operator we have to use the ghost equation of motion. To get the nilpotency off-shell, we introduce an auxiliary field $F^{a}$. So now relevant part of the modified Lagrangian is
	\begin{eqnarray}\hspace{-.5cm}
		\mathcal{L_{GF+GH}}=\frac{\xi}{2}F^{a}F^{a}+\partial^{t}F^{a}\phi^{a}+\partial^{i}F^{a}a_{i}-gf^{abc}\partial_{t}\bar{c}^{a}c^{b}\phi^{c}
		+\partial_{i}\bar{c}^{a}(\partial_{i}c^{a}-gf^{abc}c^{b}a_{i}^{c}).
	\end{eqnarray}
	Now the transformations which keep this Lagrangian invariant are given below. The transformations of gauge fields and ghost field are same as before, however the transformation of anti-ghost field is changed and written in terms $F^{a}$. The transformations are
	\begin{eqnarray}&&
		\delta\phi^{a}=\omega f^{abc}\phi^{b}c^{c},\,
		\delta a_{t}^{a}=\frac{\omega}{g}D_{t}c^{a},\,
		\delta a_{i}^{a}=\frac{\omega}{g}(D_{i}c^{a}),\non\\&&
		\delta c^{a}=-\frac{\omega}{2}f^{abc}c^{b}c^{c},\,
		\delta \bar{c}^{a}=\frac{\omega}{g}F^{a},\,
		\delta F^{a}=0.
	\end{eqnarray}
	Now doing the same analysis as before and using the Jacobi identity of structure constants, we see that these transformations are nilpotent for all the fields without using any equations of motion. The Lagrangian is invariant under these transformations without any total derivetive term.
	Now let us calculate the current for the above transformations
	\begin{eqnarray}&&
		J^{t}=-\omega f^{abc}\phi^{b}c^{c}D_{t}\phi^{a}+\frac{\omega}{g}D_{i}\phi^{a}D_{i}c^{a}+{\omega}F^{a}f^{abc}c^{b}\phi^{c}\non\\&&
		\hspace{.5cm}=-\omega f^{abc}\phi^{b}c^{c}\Pi_{\phi}^{a}+\frac{\omega}{g}\Pi_{a_{i}}^{a}D_{i}c^{a}-\frac{\omega}{g}F^{a}\Pi_{\bar{c}}^{a}.
	\end{eqnarray}
	Then the BRST charge is
	\begin{eqnarray}
		Q=\int d^{3}x J^{t}.
	\end{eqnarray}
	The Lagrangian respect another symmetry called ghost scaling symmetry. In relativistic Yang-Mills theories, this symmetry is associated with number of ghost field.
	The ghost scaling transformation is given by
	\begin{eqnarray}
		\delta c^{a}= \epsilon c^{a},\, \delta \bar{c}^{a}=-\epsilon \bar{c}^{a},
	\end{eqnarray}
	where $\epsilon$ is a constant commuting parameter. The conserved charge corresponding to this symmetry is 
	\begin{eqnarray}
		Q_{c}=\int d^{3}x J^{t}_{c}=\int d^{3}x \big(gf^{abc}\bar{c}^{a}c^{b}\phi^{c}\big)=-\int d^{3}x \bar{c}^{a}\Pi_{\bar{c}}^{a}.
	\end{eqnarray}
	The usual equal time (anti-) commutation relations are
	\begin{eqnarray}&&
		\big[\phi^{a},\Pi_{\phi}^{b}\big]=i\delta^{ab}\delta^{3}(x-y),\,
		\big[a_{i}^{a},\Pi_{a_{k}}^{b}\big]=i\delta^{ab}\delta_{ik}\delta^{3}(x-y)\non\\\non&&
		\big[a_{t}^{a},\Pi_{a_{t}}^{b}\big]=i\delta^{ab}\delta^{3}(x-y),\,
		\big[F^{a},\Pi_{F}^{b}\big]=i\delta^{ab}\delta^{3}(x-y)\\&&
		\big\{c^{a},\Pi_{c}^{b}\big\}=i\delta^{ab}\delta^{3}(x-y),\,
		\big\{\bar{c}^{a},\Pi_{\bar{c}}^{b}\big\}=i\delta^{ab}\delta^{3}(x-y)
	\end{eqnarray}
	Using these we can see that
	\begin{eqnarray}
		\big\{Q,Q\}=2Q^{2}=0,
	\end{eqnarray}
	which shows the nilpotency of the BRST operator. For the ghost operator $Q_{c}$ we have
	\begin{eqnarray}
		\big[Q_{c},Q_{c}\big]=0.
	\end{eqnarray}
	To understand the relation between $Q$ and $Q_{c}$, we will do the following analysis:
	\begin{eqnarray}&&
		(\delta_{c}\delta_{B}-\delta_{B}\delta_{c})\Phi^{a}=\delta_{B}\Phi^{a},
	\end{eqnarray}
	where $\Phi^{a}$ is any field ($\phi^{a},\,a_{t}^{a},\,a_{i}^{a},\,c^{a},\,\bar{c}^{a},\,F^{a}$). Also doing a similar computation in terms of the operators, we get
	$\big[Q,Q_{c}\big]\Phi^a = Q\Phi^a$ 
	from which we conclude that
	\begin{eqnarray}
		\big[Q,Q_{c}\big]=Q.
	\end{eqnarray}
	With many application there is a very good mathematical foundation of BRST symmetry. Using BRST symmetry in relativistic QFTs we can define equivalence class of physical states which constitute the BRST cohomology. In future, we want to do similar analysis in Galilean field theories.
	
	\subsection{Carrollian Yang-Mills  theory}\label{BRST Carrollian Electric}\label{BRSTCarrollian}
	In this section, we will examine how the BRST symmetry is manifested in the Carrollian Yang-Mills theory. Specifically, we will first investigate the realization of the BRST symmetry at the classical level for both the electric and magnetic sectors of the theory. After that, we will proceed to calculate the BRST charge. 
	
	Additionally, there is an important symmetry, denoted as $U(1)$, that exists in the ghost part of the Lagrangian, corresponding to the conservation of ghost particle number. It is important to note that the ghost field is a Grassmann variable, and therefore the algebra of charge corresponding to this $U(1)$ symmetry is anti-commuting.

	\subsubsection*{Electric Sector}
	The Lagrangian for the electric sector with gauge fixing term and ghost term is discussed in \cite{Islam:2023rnc}. The full Lagrangian for the electric sector is
	\begin{eqnarray}\label{0elecgg}
		\mathcal{L}=\frac{1}{2}E_{i}^{a(0)}E_{i}^{a(0)}-\frac{1}{2\chi}\p_{t}a_{t}^{a}\p_{t}a_{t}^{a}++\p_{t}\bar{c}^{a}D_{t}c^{a}.
	\end{eqnarray}
	Similar to the relativistic this Lagrangian also satisfy a global symmetry. The global transformations under which this Lagrangian is invariant is given by 
	\begin{eqnarray}\label{onshellbrst}
		\delta a_{t}^{a}=\frac{\omega}{g}\big(D_{t}c\big)^{a}\quad \delta a_{i}^{a}=\frac{\omega}{g}\big(D_{i}c\big)^{a}\quad \delta c^{a}=-\frac{\omega}{2}f^{abc}c^{b}c^{c},\quad \delta \bar{c}=-\frac{\omega}{g\xi}\p_{t} a_{t}^{a},
	\end{eqnarray}
	this is Carrollian BRST transformations for the electric sector.
	The action is invariant under these transformation with a total derivative term. One of the property BRST transformations is that these transformations are nilpotent. To prove that the above mentioned transformations are nilpotent we have to use Jacobi identity and equations of motion of ghost fields($\p_{t}D_{t}c^{a}=0$). The action under the transformation changes as
	\begin{eqnarray}
		\delta \mathcal{L}^{electric}=-\p_{t}\big(\frac{\omega}{g\xi}\p_{t}a_{t}^{a}D_{t}c^{a}\big)
	\end{eqnarray}
	to derive the above change we have to use ghost equations of motion. So for nilpotency of the transformations and to see the invariance of the action we need to use equations of motion for the ghost field. 
	We can also have off-shell BRST transformations which is nilpotent without using any equations of motion and the action is invariant without using any equations of motion.
	To have BRST symmetry with out using ghost equations of motion we need to introduce auxiliary field. With the auxilary field full Lagrangian is
	\begin{eqnarray}\label{fullelectric}
		\mathcal{L}^{electric}=\frac{1}{2}E_{i}^{a(0)}E_{i}^{a(0)}+\frac{\xi}{2}F^{a}F^{a}+\p_{t}F^{a}a_{t}^{a}+\p_{t}\bar{c}^{a}D_{t}c^{a}
	\end{eqnarray}
	using equation of motion of $F^{a}$ we can go back to previous Lagrangian. Now the transformations are
	\begin{eqnarray}\label{BRST trans}
		\delta a_{t}^{a}=\frac{\omega}{g}\big(D_{t}c\big)^{a}\quad \delta a_{i}^{a}=\frac{\omega}{g}\big(D_{i}c\big)^{a}\quad \delta c^{a}=-\frac{\omega}{2}f^{abc}c^{b}c^{c},\quad \delta \bar{c}=-\frac{\omega}{g}F^{a},\quad \delta F^{a}=0
	\end{eqnarray}
	The action is invariant under these transformation without using any equations of motion. These symmetry of action is off-shell symmetry. 
	
	There is another symmetry corresponding to ghost number symmetry, is global $U(1)$ symmetry. Infinitesimal form of the transformation is 
	\begin{eqnarray}\label{u1ghost}
		\delta c^{a}=\e c^{a},\quad \delta \bar{c}^{a}=-\epsilon \bar{c}
	\end{eqnarray}
	The Lagrangian \eqref{fullelectric} in invariant under this transformation. 
	
	Now to go to the  Quantum we need calculate charges corresponding to BRST symmetry and above mention $U(1)$ symmetry. After that we will see relevant commutation and anti commutation relation. Charges are  for the BRST and $U(1)$ respectively are
	\begin{eqnarray}&&
		Q_{BRST}=\int d^{3}x \big[\frac{\omega}{g}E_{i}^{a}D_{i}c^{a}-\frac{\omega}{g}F^{a}D_{t}c^{a}+\frac{\omega}{2}f^{abc}c^{b}c^{c}\p_{t}\bar{c}^{a}\big]\\&&
		Q_{U(1)}=\int d^{3}x\big[\bar{c}^{a}D_{t}c^{a}+c^{a}\p_{t}c^{a}\big]
	\end{eqnarray}
	To calculate the algebra of these charges it will convenient to write it using conjugate momentum of the fields. Then we will able to usual bracket between fields and conjugate momentum. 
	Conjugate momentum corresponding to different fields are 
	\begin{eqnarray}
		\Pi_{i}^{a}=\frac{\p\mathcal{L}}{\p(\p_{t}a_{i}^{a})}=E_{i}^{a},\, \Pi^{a}=\frac{\mathcal{L}}{\p(\p_{t}F^{a})}=a_{t}^{a},\, \Pi_{c}^{a}=-\p_{t}\bar{c}^{a},\, \Pi_{\bar{c}}^{a}=D_{t}c^{a}
	\end{eqnarray}
	Charges using these definition
	\begin{eqnarray}&&
		Q_{BRST}=\int d^{3}x \big[\frac{\omega}{g}\Pi_{i}^{a}D_{i}c^{a}-\frac{\omega}{g}F^{a}\Pi_{\bar{c}}^{a}-\frac{\omega}{2}f^{abc}c^{b}c^{c}\Pi_{c}^{a}\big]\\&&
		Q_{U(1)}=\int d^{3}x\big[\bar{c}^{a}\Pi^{a}_{\bar{c}}-c^{a}\Pi_{c}^{a}\big]
	\end{eqnarray}
	Algebra satifying these operator using field Poisson bracket
	\begin{eqnarray}
		\big[Q_{BRST},Q_{BRST}\big]=0,\quad \{Q_{U(1)},Q_{U(1)}\}=0,\quad \big[Q_{BRST},Q_{U(1)}\big]=i Q_{BRST},
	\end{eqnarray}
	We can also see these relations using \eqref{BRST trans} and \eqref{u1ghost}. From first commutation we can confirm the nilpotency of BRST symmetry. From second relation we can say that the ghost particle symmetry is abelian. Lastly from third relation we can see BRST transformation carries unit ghost particle. 
	\subsubsection*{Magnetic Sector}
	The full magnetic sector Lagrangian with gauge fixing term and ghost term is(details in \cite{Islam:2023rnc})
	\begin{eqnarray}\label{1maggg}
		\mathcal{L}=\xi_{i}^{a}E_{i}^{a}-\frac{1}{4}f_{ij}^{a}f_{ij}^{a}-\frac{1}{2\chi}\p_{i}a_{i}^{a}\p_{j}a_{j}^{a}-\p_{i}\bar{c}^{a}D_{i}c^{a}.
	\end{eqnarray}. The full Lagrangian which respect the off-shell BRST transformations is
	\begin{eqnarray}
		\mathcal{L}_{magnetic}=\xi_{i}^{a}E_{i}^{a}-\frac{1}{4}f_{ij}^{a}f_{ij}^{a}+\frac{\xi}{2}F^{a}F^{a}+\p_{i}F^{a}a_{i}^{a}+\p_{i}\bar{c}^{a}D_{i}c^{a}
	\end{eqnarray}
	Off-shell BRST transformations which keep the above Lagrangian invariant is  same as eq.\eqref{BRST trans} along with transformation of $\xi_{i}^{a}$ as $\delta\xi_{i}^{a}=\omega f^{abc}\xi_{i}^{b}c^{c}$.
	For magnetic sector BRST charge contain only spatial derivative.
	\begin{eqnarray}
		Q_{BRST}=\int d^3{x}\big[\xi_{i}^{a}D_{i}c^{a}\big]
	\end{eqnarray}
	The Lagrangian invariant under u(1)symmetry mentioned in previous section. But charge corresponding that symmetry is identically zero because in the Lagrangian there is no time derivative of ghost field. The algebra for this sector is trivially realized.

		\section{Conclusions and Discussion}\label{Conclusion}

		In this paper, we investigate the BRST symmetry of Galilean and Carrollian Yang-Mills theories, which is should be crucial to a more detailed study of gauge theories in the non-Lorentzian regime. We begin by studying the Galilean Yang-Mills theory, which is a non-relativistic theory that describes the interaction between gauge fields and matter fields in a Galilean-invariant framework. We first realized the BRST symmetry for the Galilean Yang-Mills theory and redefine its Lagrangian to make it more concrete. We then analyze the classical and quantum level of the BRST symmetry and observe that it is realized in both cases.
		
		Next, we move on to Carrollian Yang-Mills theory, which is another non-Lorentzian theory that describes the interaction between gauge fields and matter fields in a Carrollian-invariant framework. We analyze the non-trivial sectors of the theory, specifically the electric and magnetic sectors. For the magnetic sector case, the $U(1)$ charge is zero, and hence the charge algebra is trivially realized. Again, we observe the BRST symmetry at both the classical and quantum level.
		
		The study of the BRST symmetry in non-Lorentzian field theories is crucial in understanding the underlying physical properties of these theories. The BRST symmetry provides a way to fix gauge ambiguities, and its realization at the classical and quantum level is an essential aspect in the computation of physical observables.
		
		In future work, we plan to construct different sectors of the Galilean Yang-Mills Lagrangian and analyze their BRST symmetry as we discussed for Carrollian case in this paper. We also aim to extend our analysis to construct BRST cohomology for non-Lorentzian theories. This analysis will help us gain a better understanding of the fundamental symmetries of non-Lorentzian field theories and their physical properties.
		
		Our investigation of the BRST symmetry of Galilean and Carrollian Yang-Mills theories provides valuable insights into the behavior of non-Lorentzian field theories. The BRST symmetry is an essential tool in understanding the physical properties of these theories, and its realization at both the classical and quantum level is a crucial aspect in the computation of physical observables.
		
		\section*{Acknowledgments}
		
		We express our heartfelt gratitude to Arjun Bagchi for fruitful discussions, insightful suggestions, and valuable comments on this manuscript and our work. We would also like to thank Nilay Kundu, Rudranil Basu, Kedar Kolekar, Kunal Pal, and Kuntal Pal for productive discussions.

				\bibliographystyle{JHEP}
				\bibliography{reference}
			\end{document}